\documentclass[a4paper,11pt]{article}
\usepackage{pos}
\usepackage{orcidlink}
\usepackage{subcaption}

\newcommand{\orcidauthorBENNETT}{0000-0002-1678-6701}
\newcommand{\orcidauthorLUCINI}{0000-0001-8974-8266}
\newcommand{\orcidauthorPIAI}{0000-0002-2251-0111} 
\newcommand{\orcidauthorFORZANO}{0000-0003-0985-8858}
\newcommand{\orcidauthorVADACCHINO}{0000-0002-5783-5602}
\newcommand{\orcidauthorHONG}{0000-0002-3923-4184}
\newcommand{\orcidauthorLIN}{0000-0003-3743-0840}
\newcommand{\orcidauthorLEE}{0000-0002-4616-2422}
\newcommand{\orcidauthorZIERLER}{0000-0002-8670-4054}
\newcommand{\orcidauthorHSIAO}{0000-0002-8522-5190}

\title{Progress on lattice study of the chimera baryon spectrum in Sp(4) gauge theory}

\author*[a,b]{C.-J.~David~Lin\,\orcidlink{\orcidauthorLIN}}
\author[c]{Ed~Bennett\,\orcidlink{\orcidauthorBENNETT}}

\author[d]{Niccolò Forzano\,\orcidlink{\orcidauthorFORZANO}}

\author[e,f]{Deog~Ki~Hong\,\orcidlink{\orcidauthorHONG}}
\author[a,g]{Ho~Hsiao\,\orcidlink{\orcidauthorHSIAO}}
\author[h]{Jong-Wan~Lee\,\orcidlink{\orcidauthorLEE}}
\author[c,i]{Biagio~Lucini\,\orcidlink{\orcidauthorLUCINI}}
\author[d]{Maurizio~Piai\,\orcidlink{\orcidauthorPIAI}}
\author[j]{Davide~Vadacchino\,\orcidlink{\orcidauthorVADACCHINO}}
\author[d]{Fabian Zierler\,\orcidlink{\orcidauthorZIERLER}}

\affiliation[a]{Institute of Physics, National Yang Ming Chiao Tung
  University, Hsinchu 30010, Taiwan}

\affiliation[b]{Centre for High Energy Physics, Chung Yuan Christian
  University, Taoyuan 32023, Taiwan}

\affiliation[c]{Swansea Academy of Advanced Computing, Swansea
  University, Bay Campus, Swansea, SA1 8EN, UK}

\affiliation[d]{Department of Physics, Faculty  of Science and Engineering,  Swansea University, Singleton
  Park, Swansea, SA2 8PP, UK}

\affiliation[e]{Department of Physics, Pusan National University, Busan 46241, Korea}

\affiliation[f]{Extreme Physics Institute, Pusan National University, Busan 46241, Korea}

\affiliation[g]{Center for Computational Sciences, University of
  Tsukuba, Tsukuba, Ibaraki 305-8577, Japan}

\affiliation[h]{Particle Theory and Cosmology Group, Center for Theoretical Physics of the Universe, Institute for Basic Science, Daejeon, 34126, Korea}

\affiliation[i]{Department of Mathematics, Faculty  of Science and Engineering, Swansea University, Bay
  Campus, Swansea, SA1 8EN, UK}

\affiliation[j]{Centre for Mathematical Sciences, University of
  Plymouth, Plymouth, PL4 8AA, UK}

\emailAdd{dlin@nycu.edu.tw}

\abstract{Investigation of composite Higgs
  models (CHMs) is of importance in contemporary particle physics.  In this
  article, we present lattice computations of the chimera baryon masses in $Sp(4)$
  gauge theory with two and three Dirac flavours of hyperquarks (beyond the Standard Model fermions coupled to the $Sp(4)$ gauge fields) in the
  fundamental and antisymmetric representations, respectively.
 The chimera baryons are crucial for generating the Standard Model
 fermion masses through the partial compositeness mechanism in this
 gauge theory that can serve as the ultraviolet completion of the CHM with pseudo-Nambu-Goldstone bosons in the coset $SU(4)/Sp(4)$.
 Results shown here are primarily from a completed quenched computation, while
 those from our ongoing work with dynamical
 simulations are also discussed.}

\FullConference{The XVIth Quark Confinement and the Hadron Spectrum Conference (QCHSC24)\\
 19-24 August, 2024\\
 Cairns Convention Centre, Cairns, Queensland, Australia\\}

\begin{document}
\maketitle

\section{Introduction}
\label{sec:intro}
The Standard Model (SM) of particle physics has been a success in explaining all the
relevant experimental results hitherto.  There are, nevertheless, 
motivations to search for physics beyond it.  For instance, there is evidence that the
Higgs-Yukawa sector of the SM suffers from the triviality problem that
renders the cut-off scale indispensable (see, e.g.,
Refs.~\cite{Aizenman:2019yuo,Luscher:1987ay,Luscher:1987ek, Luscher:1988uq,Bulava:2012rb, Molgaard:2014mqa, Chu:2018ldw}), resulting in the unavoidable
conclusion that the SM is an effective theory, and its ultraviolet
(UV) completion has to be understood.  It is also phenomenologically incomplete, as it cannot explain the observed baryon asymmetry of the universe~\cite{Kajantie:1996mn, Laine:2012jy}, nor the origin of its dark matter and dark energy components.  

Despite the above compelling reasons to regard the SM as incomplete,
neither direct nor indirect searches at the energy and precision
frontiers have found any unambiguous evidence for
physics beyond it.   This means that the scale for new physics to
emerge can be well above a few TeV.  Therefore, any viable model
that serves as a replacement of the SM Higgs-Yukawa sector must contain a
scalar state (the Higgs boson) that is much lighter than new particles arising beyond the SM
(BSM).  This requirement makes it popular to design BSM models by
resorting to quantum field theories containing pseudo-Nambu-Goldstone bosons (PNGBs).  In
this context, composite Higgs models (CHMs) have been attracting
attention in the high energy physics
community~\cite{Panico:2015jxa, Cacciapaglia:2020kgq}.    These
CHMs can be constructed with strongly interacting gauge theories that involve confinement and spontaneous breaking of approximate, global symmetry, which
leads to the existence of naturally light bound-state Goldstone bosons.  Amongst
these confining theories~\cite{Cacciapaglia:2019bqz}, the $Sp(4)$ gauge theory with two Dirac
flavours of hyperquarks (BSM fermions coupled to the $Sp(4)$ gauge fields) in the fundamental representation is
particularly interesting~\cite{Barnard:2013zea,Ferretti:2013kya}.   Due to the pseudo-reality of this
gauge-group representation, in this model the global symmetry is
$SU(4)$, spontaneously broken to $Sp(4)$ in the limit where
hyperquarks are massless~\cite{Peskin:1980gc}.  The coset,
$SU(4)/Sp(4)$, then gives five PNGBs, with four of them
interpreted as the degrees of freedom in the SM Higgs doublet.  This
is the minimal model that allows for the extension to address the
issue of SM fermion masses, which we describe in the next paragraph.

In CHMs, the hierarchy between SM fermion masses can be explained
through the mechanism of partial compositeness~\cite{Kaplan:1991dc}.   In this article, we
use the top-quark mass for illustration.   The implementation of the
top partial compositeness is achieved in a CHM by first
identifying in the spectrum of the new, strongly coupled theory
a spin-1/2
bound state that carries the same quantum numbers as the top quark.
This bound state, one of the chimera baryons described in this document, can then mix with the
top quark and generate the mass of the latter.   To realise this
mechanism in the $Sp(4)$ gauge theory, it is convenient to introduce
three Dirac flavours of hyperquarks in the two-index antisymmetric
representation~\cite{Barnard:2013zea}.  Since this representation is real, the global
symmetry breaking pattern in this sector is $SU(6)/SO(6)$~\cite{Peskin:1980gc}.  The QCD
$SU(3)$ gauge group is then embedded in the unbroken
$SO(6)$, resulting in the fact that the antisymmetric hyperquarks
are QCD-colour charged.  Bound states composed of one antisymmetric and two
fundamental valence hyperquarks, the chimera baryons, have the right quantum numbers to be identified as top partners.

Our collaboration started the research programme of
\textit{Theoretical Explorations on the Lattice with Orthogonal and
  Symplectic groups} (TELOS), in which we employ the tool of lattice
field theory to examine various gauge theories that are interesting to the
high-energy physics community.  Investigation of the aforementioned $Sp(4)$ gauge
theory with hyperquarks in two different representations has been the
focus of our work\footnote{Lattice studies for CHMs have also been
  performed by other collaborations for the gauge groups of $SU(2)$~\cite{Arthur:2016dir,Drach:2020wux,Drach:2021uhl}
  and $SU(4)$~\cite{DeGrand:2015lna,DeGrand:2016htl,Ayyar:2017qdf,Ayyar:2018zuk,Ayyar:2018ppa,Ayyar:2018glg, Cossu:2019hse, DelDebbio:2022qgu}.}.  In the past few years, we have published several
papers in this research direction~\cite{Bennett:2017kga,
  Bennett:2019jzz, Bennett:2019cxd, Bennett:2020hqd, Bennett:2020qtj,
  Bennett:2022yfa, Bennett:2022gdz, Bennett:2022ftz, Bennett:2023wjw,
  Bennett:2023gbe, Bennett:2023mhh, Bennett:2023qwx, Bennett:2024cqv,
  Bennett:2024wda, Bennett:2024tex},  exploring measurable quantities such as the mass spectrum, decay constants, topology, and spectral densities\footnote{The $Sp(4)$ gauge theory has also been studied in the
  context of research in dark  matter~\cite{Kulkarni:2022bvh,Bennett:2023rsl,    Dengler:2024maq,Bennett:2024bhy}.}.  This article summarises the
progress of our work on the spectrum of the chimera baryons.  Most of
the results presented here are obtained in calculations performed in the
quenched approximation, as published in Ref.~\cite{Bennett:2023mhh} with
the data release in Ref.~\cite{DATA}, while some exploratory results
from our dynamical computations~\cite{dynamical} are also reported.

\section{Lattice implementation and ensemble parameters}
\label{sec:lattice}
In this work, lattice computations for $Sp(4)$ gauge theory are
carried out using the Wilson
plaquette action for the pure gauge sector.  Hyperquarks are described
by Wilson fermions.  Details of the notation can be found in Sect.~II.A
of Ref.~\cite{Bennett:2023mhh}.   As mentioned in the previous section, we
report mainly results from quenched calculations, where the fermion
determinant is set to be a constant.  Lattice parameters for the quenched ensembles used in
this work are summarised in Tab.~\ref{tab:ENS}.
\begin{table}[t]
   \caption{
  Lattice parameters and characterisation of quenched ensembles in
  this work, with
  $\beta$ being the bare inverse lattice coupling constant,  $N_{t}$ and
  $N_{s}$ being the number of temporal and spatial lattice sites, respectively.
  Here we also give the plaquette value, $\left < P \right >$, and
  the gradient-flow scale $w_0/a$~\cite{BMW:2012hcm} (determined in our previous work in Ref.~\cite{Bennett:2019cxd}),  for all the ensembles.
\label{tab:ENS}}
\begin{center}
\begin{tabular}{| c | c | c | c | c | c|}\hline\hline 
Ensemble name & $\beta$   & $N_t\times N^3_s$ & $\left < P \right >$  & $w_0/a$  \\ \hline 
QB1	    & $7.62$    & $48\times24^3$   & 0.6018898(94)	& 1.448(3)      \\ 
QB2	    & $7.7$     & $60\times48^3$    & 0.6088000(35)	& 1.6070(19)    \\ 
QB3	    & $7.85$    & $60\times48^3$   & 0.6203809(28)	& 1.944(3)      \\ 
QB4	    & $8.0$     & $60\times48^3$    & 0.6307425(27)	& 2.3149(12)    \\ 
QB5	    & $8.2$     & $60\times48^3$    & 0.6432302(25)	& 2.8812(21)    \\ \hline 
\end{tabular}
\end{center}
\end{table}
We also report exploratory results from our fully dynamical
simulations, obtained using the ensemble M2 with details given in Tab.~1 of Ref.~\cite{Bennett:2024wda}.

We investigate the chimera baryon states sourced by the following
operators (where $Q$ and $\Psi$ are the fundamental and antisymmetric hyperquark fields, respectively)
\begin{eqnarray}
 {\mathcal{O}}_{\rho}^{ijk,5} &\equiv& Q^{i\,a}_{\alpha} (C\gamma^{5})_{\alpha\beta} Q^{j\,b}_{\beta} \Omega^{ad}\Omega^{bc} \Psi^{k\,cd}_{\rho} \, , \label{eq:chim_bar_src} \\ 
 {\mathcal{O}}_{\rho}^{ijk,\mu} &\equiv& Q^{i\,a}_{\alpha} (C\gamma^{\mu})_{\alpha\beta} Q^{j\,b}_{\,\beta} \Omega^{ad}\Omega^{bc} \Psi^{k\,cd}_{\rho} \, ,\label{eq:chim_bar_src_mu}
\end{eqnarray}
with $\alpha, \beta$, $\rho$ being spinor indices, $i,j,k$ indicating
the flavour quantum number, and $a,b,c,d$ being hypercolour indices.
In the above expressions, 
$\gamma^5$ and $\gamma^{\mu}$ are Dirac matrices,
with $C$ being the matrix for the operation of charge
conjugation. Furthermore, $\Omega$ is the symplectic matrix,
\begin{equation}
\label{eq:Omega_def}
\Omega \equiv \left ( 
\begin{array}{cccc}
 0 & 0 & 1 & 0\\
 0 & 0 & 0 & 1\\
 -1 & 0 & 0 & 0\\
 0 & -1 & 0 & 0\\
\end{array}
\right ) \, ,
 \end{equation}
in the hypercolour space.  The operator in
Eq.~(\ref{eq:chim_bar_src}) sources spin-1/2 states.  On the other
hand, ${\mathcal{O}}_{\rho}^{ijk,\mu}$ in Eq.~(\ref{eq:chim_bar_src_mu})
overlaps with both spin-1/2 and 3/2 states, which can be distinguished
using the projectors,
\begin{equation}
  \label{eq:spin_projectors}
  P_{1/2}^{\mu\nu} = \frac{1}{3} \gamma^{\mu} \gamma^{\nu} \, , \, \,
  P_{3/2}^{\mu\nu} =  \delta^{\mu\nu}- \frac{1}{3} \gamma^{\mu}
  \gamma^{\nu} \, .
\end{equation}
Drawing analogy with QCD baryonic states, the three lowest-lying states sourced by the interpolators in
Eqs.~(\ref{eq:chim_bar_src}) and~(\ref{eq:chim_bar_src_mu}) are called
$\Lambda_{\mathrm{CB}}$,  $\Sigma_{\mathrm{CB}}$ and
$\Sigma^{\ast}_{\mathrm{CB}}$.  Their properties are summarised in
Tab.~\ref{tab:interpolators}.  Amongst these three chimera baryons,
$\Lambda_{\mathrm{CB}}$ and $\Sigma_{\mathrm{CB}}$ are candidates of the
top partner~\cite{Banerjee:2022izw}.  Note that in this article we discuss
only flavour off-diagonal chimera baryon states, and will not write
the flavour indices explicitly below.  
\begin{table}[t]
   \caption{
 Properties of the three lowest-lying flavoured  chimera baryon states sourced by the interpolating operators in
Eqs.~(\ref{eq:chim_bar_src}) and (\ref{eq:chim_bar_src_mu}) with
flavour indices omitted.  In addition to the spin quantum number, here
we also list the representations of the unbroken global symmetry
groups, $Sp(4)$ and $SO(6)$.  
\label{tab:interpolators}}
\begin{center}
\begin{tabular}{| c | c | c | c | c | c|}\hline\hline 
Chimera baryon & Interpolator   & $J$ & $Sp(4)$ repn & $SO(6)$ repn &
                                                                      QCD analogy \\ \hline 
$\Lambda_{\mathrm{CB}}$        &  $ {\mathcal{O}}_{\rho}^{5}$  & 1/2 & 5   & 6 & $\Lambda$ (1116)   \\
$\Sigma_{\mathrm{CB}}$           &   ${\mathcal{O}}_{\rho}^{\mu}$ & 1/2 & 10 & 6 & $\Sigma$ (1193)   \\
$\Sigma^{\ast}_{\mathrm{CB}}$ &   ${\mathcal{O}}_{\rho}^{\mu}$  & 3/2 & 10 & 6 & $\Sigma^{\ast}$ (1379)   \\
\hline 
\end{tabular}
\end{center}
\end{table}
We then compute the correlators for extracting masses of the
above-mentioned chimera baryons.  For $\Lambda_{\mathrm{CB}}$, it is
\begin{equation}
\label{eq:Lambda_corr}
 C_{\Lambda_{\mathrm{CB}}}(t) = \sum_{\vec{x}} \langle
 {\mathcal{O}}_{\rho}^{5} (x) \overline{ {\mathcal{O}}_{\sigma}^{5}} (0)
 \rangle \, .
\end{equation}
As for $\Sigma_{\mathrm{CB}}$ and
$\Sigma^{\ast}_{\mathrm{CB}}$, we have
\begin{equation}
  \label{eq:Sigma_spin_proj_corr}
  C_{\Sigma_{\mathrm{CB}}} (t) = P_{1/2}^{\mu\nu} C^{\mu\nu}(t) \, ,
  \, \, C_{\Sigma^{\ast}_{\mathrm{CB}}} (t) = P_{3/2}^{\mu\nu}
  C^{\mu\nu}(t) \, ,
\end{equation}
with
\begin{equation}
  \label{eq:Sigma_corr_before_proj}
  C^{\mu\nu}(t) = \sum_{\vec{x}} \langle {\mathcal{O}}_{\rho}^{\mu} (x) \overline{ {\mathcal{O}}_{\sigma}^{\nu}} (0)\rangle \, .
\end{equation}
Note that we do not write explicitly the dependence on the spinor indices, $\rho$ and $\sigma$, in $C_{\Lambda_{\mathrm{CB}}}(t)$, $ C_{\Sigma_{\mathrm{CB}}} (t)$, $C_{\Sigma^{\ast}_{\mathrm{CB}}} (t)$, and $C^{\mu\nu}(t)$. In the following, we will use the symbol, $C_{\mathrm{CB}} (t)$, to
denote a generic chimera-baryon correlator in Eqs.~(\ref{eq:Lambda_corr}) and (\ref{eq:Sigma_spin_proj_corr}).

It should be noted that the interpolating operators in
Eqs.~(\ref{eq:chim_bar_src}) and (\ref{eq:chim_bar_src_mu}) source
both even and odd parity states.  In fact, without an actual lattice
numerical calculation, it is not easy to confirm the parity of the
lowest-lying states appearing in the correlators in
Eqs.~(\ref{eq:Lambda_corr}) and (\ref{eq:Sigma_spin_proj_corr}).  In
other words, it requires a lattice calculation to determine the parity
of $\Lambda_{\mathrm{CB}}$, $\Sigma_{\mathrm{CB}}$ and
$\Sigma^{\ast}_{\mathrm{CB}}$.  This will be discussed in the next section.

\section{Analysis and numerical results}
\label{sec:results}
Since the interpolators in Eqs.~(\ref{eq:chim_bar_src}) and (\ref{eq:chim_bar_src_mu}) source both parity-even and -odd
states, at large Euclidean time, a chimera-baryon correlator is expected to exhibit the
behaviour,
\begin{equation}
  \label{eq:c_CB_eo}
C_{\rm CB}(t) \xrightarrow{0 \ll t \ll T} P_{+} \left[ c_+e^{-m^+ t} - c_- e^{-m^-(T-t)} \right] + P_{-} \left[ c_- e^{-m^- t} - c_+ e^{-m^+ (T-t)} \right]\,,
\end{equation}
with $T$ being the temporal lattice extent, and $P_{\pm}$ being the
parity projectors,
\begin{equation}
  \label{eq:parity_projectors}
  P_{\pm} = \frac{1\pm \gamma_{0}}{2} \, .
\end{equation}
The quantities $m^\pm$ and $c_\pm$ are masses and the
baryon-to-vacuum transition amplitudes of the corresponding parity states.
To proceed, we compute the ``parity-projected'' correlator,
\begin{equation}
  \label{eq:parity_projected_corr}
\overline{C}_{\rm CB}^\pm (t) = \frac{P_{\pm}C_{\rm CB} (t) - P_{\mp}C_{\rm CB} (T-t)}{2} \, .
\end{equation}
It can be easily shown that the large-Euclidean-time behaviour of $\overline{C}_{\rm CB}^\pm (t)$ is
\begin{equation}
\overline{C}^{\pm}_{\rm CB} (t) \longrightarrow c_{\pm} e^{-m^\pm t} -c_{\mp} e^{-m^\mp (T-t)}\, .
\label{eq:corr_eo}
\end{equation}

\begin{figure}[t]
\centering
    \includegraphics[width=7.5cm, height=6cm]{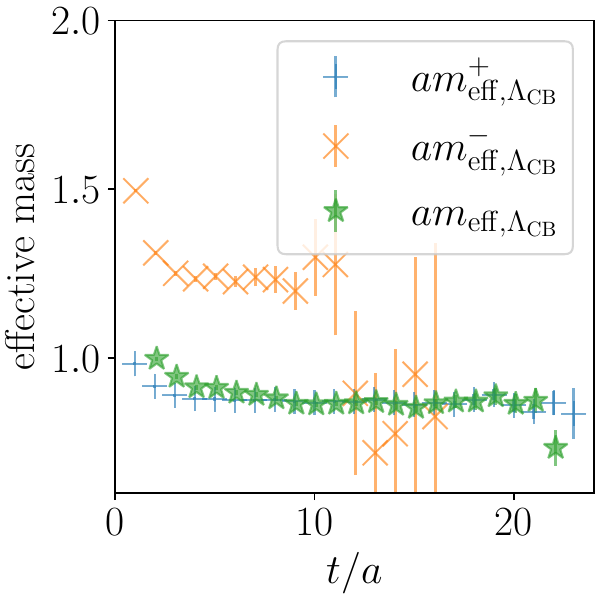}
\caption{The $\Lambda_{\mathrm{CB}}$ baryon effective masses computed from parity-projected and unprojected correlators, on ensemble QB1 in Tab.~\ref{tab:ENS} with the bare fundamental and antisymmetric valence hyperquark masses (in units of the lattice spacing, $a$) $am_0^{(f)}=-0.77$ and  $am_0^{(as)}=-1.1$.
Original plot is in Fig.~1(b) of Ref.~\cite{Bennett:2023mhh}.}
\label{fig:projection}
\end{figure}

Working in the regime $0 \le t < T/2$, one can define the effective
masses (in lattice units) for the parity-even and -odd states,
\begin{equation}
am^{\pm}_{\rm eff, CB}(t) = \ln{\left [ \frac{\overline{C}^{\pm}_{\rm CB}(t)}{\overline{C}^{\pm}_{\rm CB}(t+1)} \right] }\, .
\label{eq:meff_cb}
\end{equation}
Similarly, one can also calculate the effective mass using the
unprojected correlator in Eq.~(\ref{eq:c_CB_eo}),
\begin{equation}
am_{\rm eff, CB}(t) = \ln{\left [ \frac{C_{\rm CB}(t)}{C_{\rm CB}(t+1)} \right] }\, .
\label{eq:meff_cb_unproj}
\end{equation}
Upon examination of the effective masses in Eq.~(\ref{eq:meff_cb}), one can determine the parity of the ground-state chimera baryons.  Furthermore, the asymptotic behaviour of the effective mass computed through the unprojected correlator can serve as a crosscheck.  One example is illustrated in Fig.~\ref{fig:projection}, where we display these effective masses for states sourced by $\mathcal{O}^{5}_{\rho}$.  It is clear that the lightest chimera baryon in this channel ($\Lambda_{\rm{CB}}$) is of even parity.  We find the same conclusion for $\Sigma_{\mathrm{CB}}$ and $\Sigma_{\mathrm{CB}}^{\ast}$.
%
%
%

In this work, we extract the chimera baryon masses, $m_{\mathrm{CB}}$, from correlators at various fundamental and antisymmetric bare hyperquark masses, $m^{(f)}$ and $m^{(as)}$, on the ensembles listed in Tab~\ref{tab:ENS}.  Using formulae inspired by heavy baryon chiral perturbation theory (HBChPT)~\cite{Jenkins:1990jv,Bernard:1995dp}, with the inclusion of symmetry-breaking effects from lattice artefacts to $\mathcal{O}(a)$~\cite{Beane:2003xv}, this enables us to investigate the hyperquark-mass dependence of $m_{\mathrm{CB}}$.  It also allows for the extrapolation to the continuum limit.  We work at the order of the square of the hyperquark masses, {\it i.e.}, the fourth power of the fundamental and antisymmetric pseudoscalar-meson masses, $m_{\mathrm{PS}}$ and $m_{\mathrm{ps}}$.  As explained in detail in Ref.~\cite{Bennett:2023mhh}, it is challenging to fit our data to the expansion at this order, and we explore a strategy to systematically exclude data points and terms in the fit formulae.  This strategy results in 1315 different analysis procedures.  An Akaike information criterion (AIC)~\cite{Akaike:1998zah} is then employed to select the optimal procedure.  Using this method, we find that for $m_{\Lambda_{\mathrm{CB}}}$ and $m_{\Sigma_{\mathrm{CB}}}$, the best fit function is 
\begin{eqnarray}
    \hat{m}_{\textrm{CB}} &=& \hat{m}_{\rm CB}^\chi + F_2 \hat{m}_{\rm PS}^2 + A_2 \hat{m}_{\rm ps}^2 + L_1 \hat{a}  \nonumber \\ 
    && \hspace{0.77cm}+ F_3 \hat{m}_{\rm PS}^3 + A_3 \hat{m}_{\rm ps}^3 + L_{2F} \hat{m}_{\rm PS}^2 \hat{a}+ L_{2A}  \hat{m}_{\rm ps}^2 \hat{a} \nonumber \\
&& \hspace{0.77cm}+ C_{4}  \hat{m}_{\rm PS}^2  \hat{m}_{\rm ps}^2\,,
\label{eq:fitting_func}
\end{eqnarray}
where the hatted symbols denote dimensionful quantities expressed in units of the gradient-flow scale $w_{0}$.  In the above equation, $\hat{m}^{\chi}_{\mathrm{CB}}$, $F_{i}$, $A_{i}$, $L_{i}$ and $C_{4}$ are the "low-energy constants" (LECs), to be determined by the fit. Regarding the mass of $\Sigma^{\ast}_{\mathrm{CB}}$, the same fit function without the $C_{4}  \hat{m}_{\rm PS}^2  \hat{m}_{\rm ps}^2$ term is optimal. 

\begin{figure}
	\centering	\includegraphics[width=\textwidth]{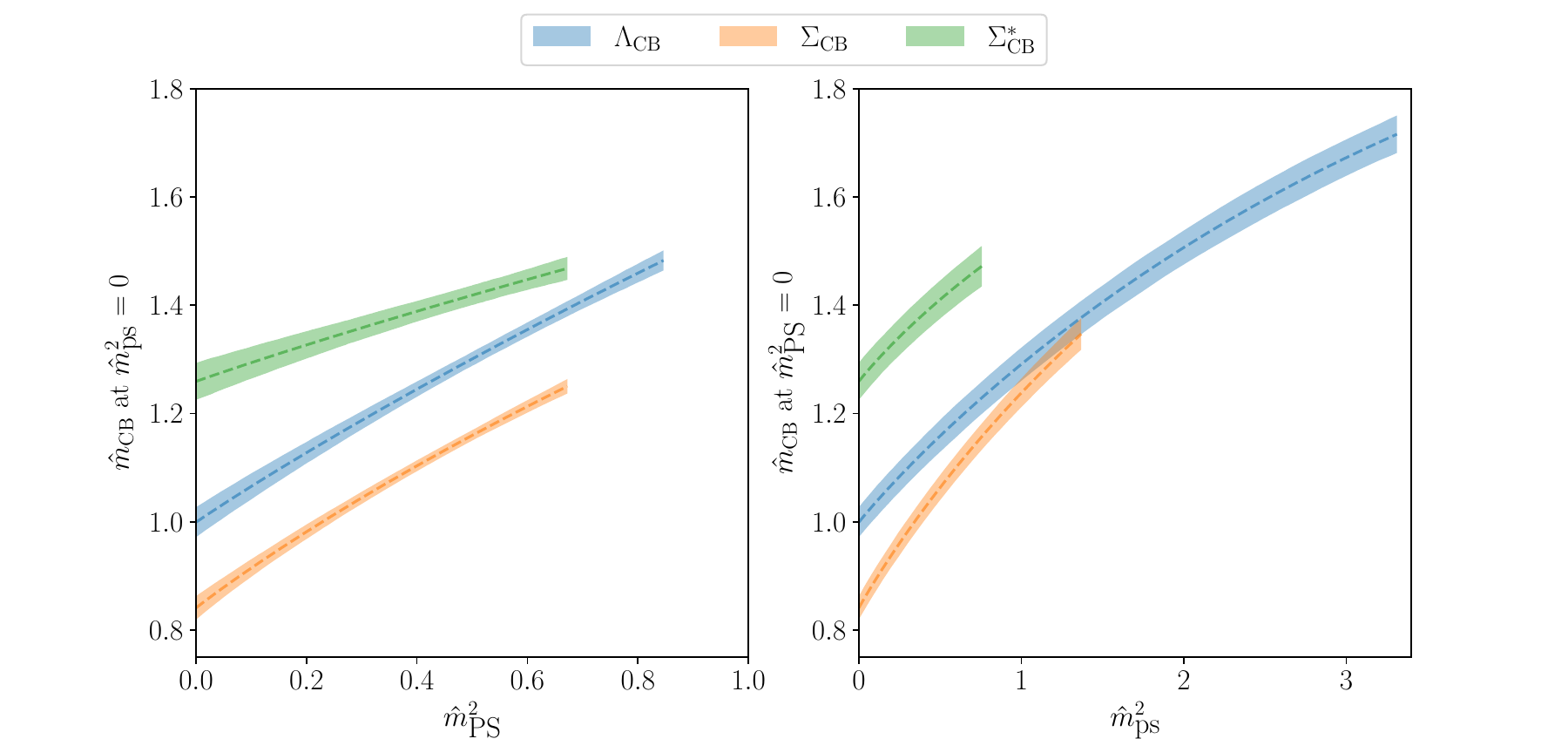}
	\caption{Mass dependence of the chimera baryons, in the continuum limit, as a function of the mass squared of the pseudoscalars, $\hat{m}^{2}_{\rm PS}$ (left) and $\hat{m}^{2}_{\rm ps}$ (right).  Original plots are in Fig.~12 of Ref.~\cite{Bennett:2023mhh}.
    }	\label{fig:CB_extrapolation}
\end{figure}

\begin{figure}
	\centering	\includegraphics[width=\textwidth]{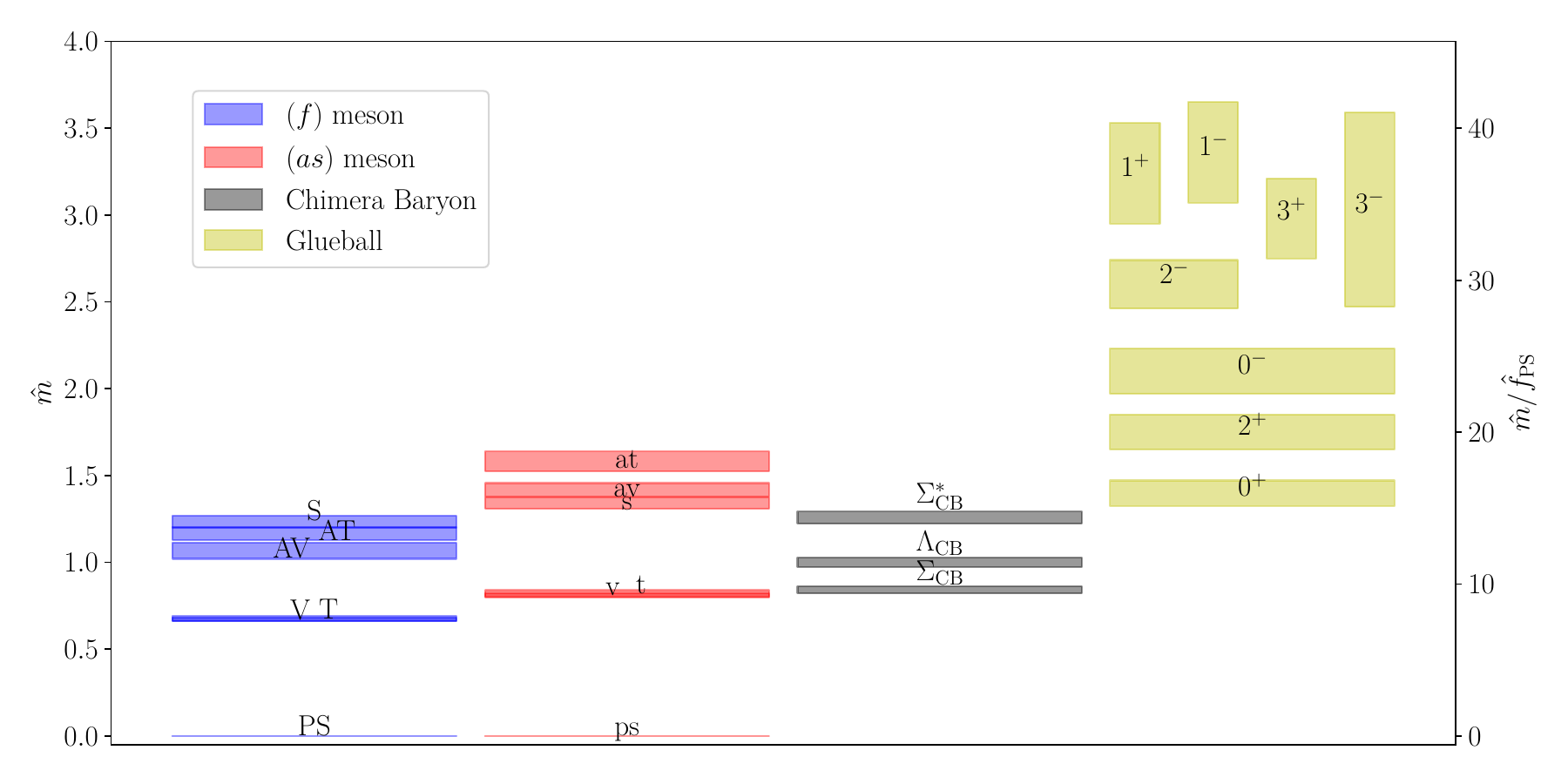}
	\caption{Meson~\cite{Bennett:2019cxd}, glueball~\cite{Bennett:2020qtj} and chimera baryon masses in the quenched $Sp(4)$ gauge theory extrapolated to the limit of zero hyperquark masses. Glueball states are represented by the spin and parity quantum numbers with the notation $J^{P}$. In this plot, PS (ps), V (v), T(t), AV (av), AT (at) and S (s) denote the pseudoscalar, vector, tensor, axial-vector, axial-tensor and scalar mesons (off-diagonal in hyperquark flavours) made of fundamental (antisymmetric) hyperquarks. The symbol $f_{\mathrm{PS}}$ means the decay constant of the PS meson in the vanishing-hyperquark-mass limit.  Original plot is in Fig.~13 of Ref.~\cite{Bennett:2023mhh}.
    }
	\label{fig:quench_spec}
\end{figure}

Figure~\ref{fig:CB_extrapolation} illustrates the results of the above HBChPT-inspired analysis strategy.
Plots in this figure are prepared with $L{_1}=L_{2F}=L_{2A}=0$.  That is, they show the results in the continuum limit.  The plot on the left-hand side exhibits results with $\hat{m}_{ps}=0$, while that on the right-hand side is for $\hat{m}_{PS}=0$.  From these plots, it is clear that $m_{\Sigma^{\ast}_{\mathrm{CB}}} > m_{\Lambda_{\mathrm{CB}}} \ge m_{\Sigma_{\mathrm{CB}}}$ in the range of the hyperquark masses that we explore. Note that this hierarchy is different from that of the analogous QCD baryons in Tab.~\ref{tab:interpolators}.  In this work, we also extrapolate the chimera baryon masses to the limit where $m_{\mathrm{PS}}$ and $m_{\mathrm{ps}}$ are both vanishing.  In Fig.~\ref{fig:quench_spec} we demonstrate these results together with the quenched meson and glueball spectra published in Refs.~\cite{Bennett:2019cxd,Bennett:2020qtj}.

\begin{figure}
\centering
   \includegraphics[width=\textwidth]{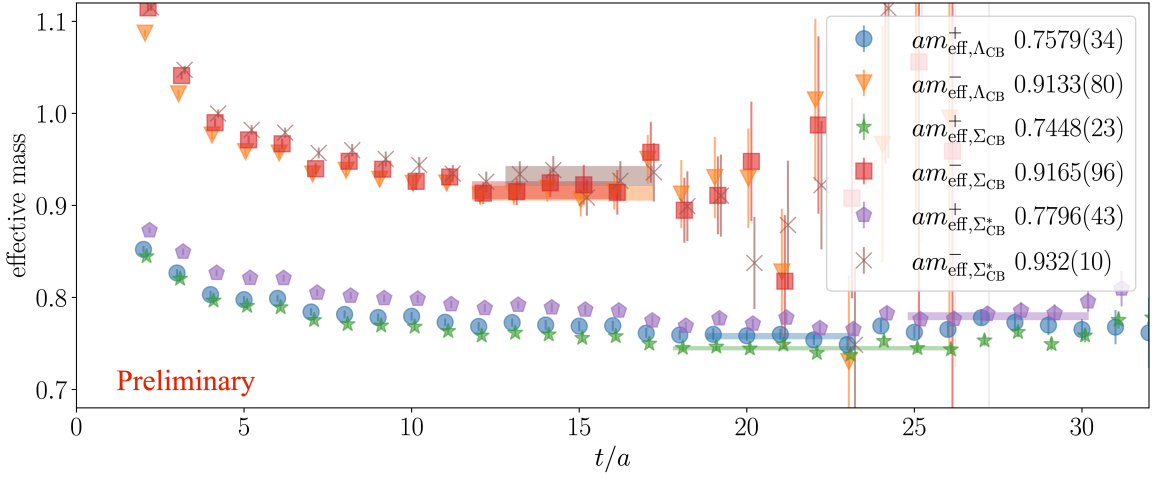}
 \caption{
Ground-state chimera baryon effective masses computed on the dynamical ensemble M2 (details given in Tab.~1 of Ref.~\cite{Bennett:2024wda}).  The coloured bands illustrate the fits for the masses, with the horizontal width representing the fit range, and the height indicating the statistical error.
Here we display results for both parity-even and -odd states.}
\label{fig:MIX_chimera_M2}
\end{figure}

\begin{figure}
	\centering	\includegraphics[width=\textwidth]{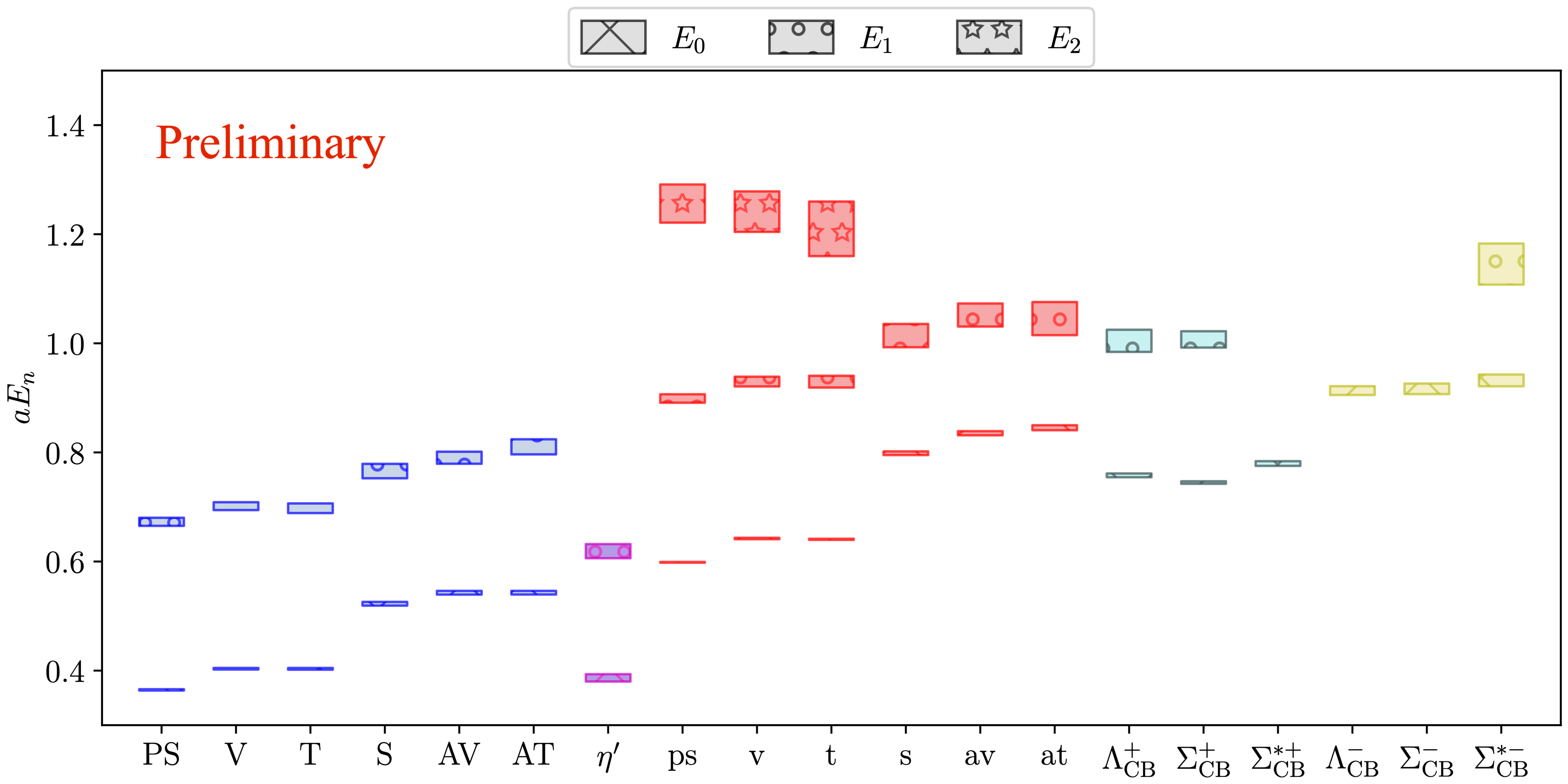}
	\caption{Masses of mesons and chimera baryons extracted from the dynamical ensemble M2 in Tab.~1 of Ref.~\cite{Bennett:2024wda}).  The symbol $E_{0}$ means the ground-state mass, while $E_{1}$ and $E_{2}$ indicate the first and the second excited state masses, respectively.  A GEVP method is used in the analysis.  Pseudoscalar, vector, tensor, axial-vector, axial-tensor and scalar mesons (off-diagonal in hyperquark flavours) made of fundamental (antisymmetric) hyperquarks are denoted by PS (ps), V (v), T(t), AV (av), AT (at) and S (s).  Description of the analysis procedure for obtaining these meson masses is reported in Ref.~\cite{Hsiao:2024tjf}).  Furthermore, results for the flavour-siglet mesons (details in Ref.~\cite{Bennett:2024wda}), represented generically by $\eta^{\prime}$, are also ploted here.
    }
	\label{fig:dyn_full_spec_M2}
\end{figure}

In addition to the quenched spectrum, here we also present preliminary results of chimera baryon masses in our dynamical simulations with 2 fundamental and 3 antisymmetric Dirac flavours of hyperquarks.  An effective mass plot using data from our ensemble M2 (details given in Tab.~1 of Ref.~\cite{Bennett:2024wda}) is shown in Fig.~\ref{fig:MIX_chimera_M2}, where we also display results for parity-odd states. 
These chimera baryon masses are then exhibited together with meson masses obtained on the same ensemble in Fig.~\ref{fig:dyn_full_spec_M2}.  
Results for the excited-state mesons in this figure are extracted from an analysis using the the method of generalised eigenvalue problem (GEVP)~\cite{Luscher:1990ck}.



%
\section{Conclusion and outlook}
\label{sec:conclusion}
The chimera baryons in CHMs play an essential role in the implementation of the partial compositeness mechanism that facilitates the generation of the SM fermion masses, hence understanding their properties is of interest in the particle physics community.  In this article, we present lattice calculations of the spectrum of low-lying chimera baryons in the $Sp(4)$ gauge theory coupled to two and three Dirac flavours of hyperquarks transforming in the fundamental and antisymmetric representations of the gauge group, respectively. 
Our calculation shows that candidates of the top partner, which are identified with the $\Lambda_{\mathrm{CB}}$ or $\Sigma_{\mathrm{CB}}$, as well as the lightest spin-3/2 state, $\Sigma^{\ast}_{\mathrm{CB}}$, are parity-even.  In this article, a completed quenched computation for $m_{\Lambda_{\mathrm{CB}}}$, $m_{\Sigma_{\mathrm{CB}}}$ and $m_{\Sigma^{\ast}_{\mathrm{CB}}}$ with extrapolations to the continuum limit is presented.  We also discuss preliminary results of our ongoing work on calculations of these chimera baryon masses with dynamical hyperquarks.  

In the future, we will also compute matrix elements entering the transitions between the top-partner candidates and the vacuum.  These matrix elements are crucial input for determining the SM fermion masses from the partial compositeness mechanism, since they govern the mixing strength between SM fermions and their partner chimera baryons.  

\begin{acknowledgments}

EB, BL, MP, FZ acknowledge the STFC DTP Grant No.~ST/X000648/1. EB and BL are grateful for the support of their work from the EPSRC ExCALIBUR programme ExaTEPP (project EP/ X017168/1).  
EB also acknowledges the STFC Research Software Engineering Fellowship EP/V052489/1.
The work of NF is supported by the STFC Doctoral Training Grant No. ST/X508834/1.
DKH acknowledges support from Basic Science Research Program through the National Research Foundation of Korea (NRF) funded by the Ministry of Education (NRF-2017R1D1A1B06033701), as well as the NRF grant MSIT 2021R1A4A5031460 from the Korean government.
JWL acknowledges support from Institute for Basic Science through the project code IBS-R018-D1. 
HH and CJDL acknowledges support from NSTC Taiwan, through grant number 112-2112-M-A49-021-MY3.
CJDL also receives support from two other NSTC grants, 112-2639-M-002-006-ASP and 113-2119-M-007-013.  DV acknowledges support from STFC through Consolidated Grant No.~ST/X000680/1.
BL and MP are also funded by STFC through Consolidated Grant No.~ST/T000813/1, as well as by the European Research Council (ERC) through Horizon 2020 research and innovation program of the European Union, through Grant Agreement No.~813942.

This work used the DiRAC Data Intensive service CSD3 at the University of Cambridge, managed by the University of Cambridge University Information Services; the DiRAC Data Intensive service DIaL at the University of Leicester, managed by the University of Leicester Research Computing Service; and the DiRAC Extreme Scaling service Tursa at the University of Edinburgh, managed by the Edinburgh Parallel Computing Centre; all on behalf of the STFC DiRAC HPC Facility (www.dirac.ac.uk), and each of which was funded by BEIS, UKRI and STFC capital funding and STFC operations grants. DiRAC is part of the UKRI Digital Research Infrastructure. We acknowledge the support of the Supercomputing Wales programme, which was part-funded by the European Regional Development Fund via Welsh Government.

{\bf Open Access Statement}---For the purpose of open access, the authors have applied a Creative Commons 
Attribution (CC BY) licence to any Author Accepted Manuscript version arising.

{\bf Research Data Access Statement}---The results reported here are based on preliminary analysis.
Further analysis and the data generated for this manuscript will be released together with an upcoming publication. Alternatively, preliminary data and code can be obtained from the authors upon request.

\end{acknowledgments}
\bibliographystyle{apsrev.bst}
\bibliography{refs.bib}

\end{document}